\documentclass{emulateapj}
\pdfoutput=1
\usepackage{amsmath,amstext}
\usepackage{lineno}
\usepackage{graphicx}
\usepackage{amssymb}
\usepackage{epstopdf}
\usepackage{aas_macros}
\usepackage{booktabs}
\usepackage{hyperref}
\usepackage{rotating}
\usepackage{mathtools}

\usepackage{color}

\newcommand{\cg}[1]{\textcolor{black}{ #1}}

\def\msun{M_\odot}

\begin{document}

\title{Distance measures in gravitational-wave astrophysics and cosmology}
\author{Hsin-Yu Chen\altaffilmark{1}, Daniel E. Holz\altaffilmark{2}, John Miller\altaffilmark{3}, {Matthew Evans\altaffilmark{3}, Salvatore Vitale\altaffilmark{3}}, Jolien Creighton\altaffilmark{4}}

\altaffiltext{1}{Black Hole Initiative, Harvard University, Cambridge, Massachusetts 02138, USA}
\altaffiltext{2}{University of Chicago, Chicago, Illinois 60637, USA}
\altaffiltext{3}{LIGO, Massachusetts Institute of Technology, Cambridge, Massachusetts 02139, USA}
\altaffiltext{4}{University of Wisconsin-Milwaukee, Milwaukee, Wisconsin 53201, USA}

\begin{abstract}
We present quantities which characterize the sensitivity of gravitational-wave observatories to sources at cosmological distances.
In particular, we introduce and generalize the horizon, range, response, and reach distances.
These quantities incorporate a number of important effects,
including cosmologically well-defined distances and volumes, cosmological redshift, cosmological time dilation,
and rate density evolution. In addition, these quantities incorporate unique aspects of gravitational wave detectors,
such as the variable sky sensitivity of the detectors and the scaling of the sensitivity with inverse distance.
An online calculator (\url{https://users.rcc.uchicago.edu/~dholz/gwc/})
and python notebook (\url{https://github.com/hsinyuc/distancetool}) to determine GW distances are available.
We provide answers to the question: {``How far can gravitational-wave detectors hear?''}
\end{abstract}

\maketitle

\section{Introduction}
\label{sec:1}

Advanced LIGO-Virgo has officially ushered in the era of gravitational-wave (GW)
astrophysics. The first detections have included binary black hole systems well into the Hubble
flow where cosmological effects start to become important; for example, the
redshift of GW170729 is {$z \sim 0.49$}
~\citep{LIGOScientific:2018mvr}. As
GW detectors improve, and as the network of GW detectors is expanded{~\citep{Hild:2010id,2012CQGra..29l4013S,osd,2016PhRvL.116m1103A}}, we
expect to detect binaries to ever greater distances. With this in mind, in what
follows we present a number of quantities to summarize the sensitivity of
detectors taking into account cosmological effects such as time dilation and
cosmological volume.

Furthermore, there are some characteristics of GW astronomy that are fundamentally different from ``traditional'' electromagnetic
(EM) astronomy, and this means that quantities used to summarize EM telescopes need to be adjusted for
GW telescopes~\footnote{In many ways GW astronomy is closer to radio astronomy, where the signals are coherent and beam patterns must
be incorporated. See the discussion below.}. Quantities such as magnitude limit, sky brightness, B-band luminosity, and Vega magnitudes need to be replaced.

One particularly important distinction between optical and GW telescopes is their differing sky response. 
GW telescopes are sensitive to sources on the entire sky, although the sensitivity varies greatly depending on the 
particular sky location. The average distance {to which a GW telescope can detect a given source}
varies greatly depending on where the source is on the sky 
{\em relative to the detector}\/ (i.e., as measured in the \cg{detector frame}, not a fixed location on the sky). 
The quantities we propose below take this antenna pattern sensitivity into account. 
\cg{In Section~\ref{sec:def}, we provide the definitions of the quantities and report the values of these quantities with two different sensitivity curves. 
In Section~\ref{sec:diff}, we compare the differences of these quantities. In Section~\ref{sec:howto}, we further explain why cosmology 
plays a role in the definition of the distance measure. We summarize in Section~\ref{sec:summary}.}

\section{Distance measures}\label{sec:def}
The sensitivity of a GW detector is a function of two factors: the properties of
the detector and the properties of the source of interest. For any fixed
detector noise curve (e.g., LIGO O3) and any fixed binary coalescence system
(e.g., 30--$30\,\msun$ binary black holes), we are interested in summarizing the sensitivity of that detector to
that given source. In particular, some quantities of interest include:
\begin{itemize}
\item {\bf Horizon distance, $d^{\rm h}$}: The farthest luminosity distance the given source could ever be detected above threshold
(i.e., at optimal sky location and binary inclination/orientation). Throughout this paper we assume the detection threshold is approximated by a matched-filter
signal-to-noise ratio ($\mbox{SNR}$), $\rho_{\rm th}$, of 8{{~\citep{thorne300,2012PhRvD..85l2006A}}}. See below for more detail.

\item {\bf Redshifted Volume, $V_z$}: The spacetime volume surveyed per unit detector time, in
  units of Mpc$^3$. This is the comoving volume (see,
    e.g.,~\cite{1999astro.ph..5116H}), with the addition of a $(1+z)$ factor to
    account for time dilation.
  If you multiply $V_z$ by the constant comoving source-frame rate
    density, you get the detection rate.
  This quantity is sky-averaged and inclination/orientation averaged.
  In detail:
\begin{multline}\label{eq:vol1}
\hspace{0.5cm} V_z = \\
  \frac{\int_{D_c<d^{\rm h}} \frac{D_c^2}{1+z(D_c)}dD_c\,d\Omega\,\sin\iota\,d\iota\, d\psi}{\int \sin\iota\,d\iota\, d\psi},
  \end{multline}
where $D_c$ is the comoving distance and $\Omega$ is the solid angle on the sky.
$d^{\rm h}(\theta,\phi,\psi,\iota)$ is the comoving distance for which $\mbox{SNR}=\rho_{\rm th}$ for a binary with 
inclination $\iota${~\footnote{The inclination is the angle between the binary rotational axis and the line-of-sight direction.}},
orientation $\psi$, and along the sky direction $(\theta,\phi)$.

\item {\bf Range distance, $R$}: The distance for which $4/3\pi {R}^3 = {V_z}$, where
    $V_z$ is defined above.
This is the radius of a Euclidean sphere which would contain the same volume as the true redshifted volume.
At low redshift ($z\lesssim1$) this quantity is well approximated by the horizon distance divided by 2.264{~\citep{1993PhRvD..47.2198F}} and 
has historically been called the {``sensemon distance''}; see Sec.~\ref{sec:euclid} for more detail.

\item{\bf Response distance, $d^{\rm p}_x$}: The luminosity distance {{\em at}} which $x\%$ of the sources would be detected,
  for sources placed isotropically on the sky with random inclinations/orientations,
  {\em but with all sources placed at exactly this distance}. Note that $d^p_0$ corresponds to the horizon distance. 
  A binary at distance $d^{\rm p}_x$ would have a {maximum possible} SNR of $\rho$, and that $\rho$ would satisfy:
  {{$P(\rho_{\rm th}/\rho)=x$}}, where $P$ is the cumulative antenna pattern function (see Sec.~\ref{sec:antenna}).

\item {\bf Reach distance, $d^{\rm r}_x$}: 
The luminosity distance {{\em within}} which $x\%$ of {the total detections would take place.}
$d^{\rm r}_{50}$ corresponds to the median distance of the detected
  population of sources, and $d^p_{100}$ corresponds to the horizon distance.
  The redshifted volume out to $d^{\rm r}_x$, divided by the total redshifted volume, 
  $V_z$, is given by $x\%$. In more detail, we calculate the redshifted volume using Eq.~\ref{eq:vol1}, but with the limits 
  of the integration given by $\min(d^{\rm r}_x, d^{\rm h})$ instead of $d^{\rm h}$ (where $d^{\rm r}_x$ here is in comoving distance). If we divide this by the total redshifted volume (Eq.~\ref{eq:vol1} with $d^{\rm h}$ in the limit), we find a ratio of $x\%$.

\item {\bf SFR Reach distance, $d^{\rm SFR}_x$}: The same as $d^{\rm r}_x$, but we now scale the
  source frame rate density by the star formation rate \cg{(SFR)}. This is a very rough
  approximation to the effect of rate evolution on the detected sample.
  This is equivalent to the expression for $d^{\rm r}_x$, but with the volume integrals in $V_z$ weighted by
  an additional factor of the star formation rate.

\item{\bf Average distance, ${\bar d}$}: The average luminosity distance of the detected sample.
This is to be compared with the median luminosity distance of the detected sample, given by $d^{\rm r}_{50}$.
This is the same as the volume integrals weighted by the luminosity distance and
divide by the total redshifted volume.

\item {\bf SFR Average distance, ${\bar d}^{\rm SFR}$}: The same as the
{Average distance}, but we now additionally scale the source-frame rate  density by the star formation rate.
This is equivalent to the expression for ${\bar d}$ with an additional factor of the star formation rate in the volume integral in
both the numerator and the denominator.

\end{itemize}

{An online calculator to determine these distance measures for a range of sources and detector noise curves is available at
\url{https://users.rcc.uchicago.edu/~dholz/gwc/}. Python notebooks are also provided at
\url{https://github.com/hsinyuc/distancetool}, which provide additional details for how to calculate these expressions.}

\begin{figure*}
\centering
\includegraphics[width=\textwidth]{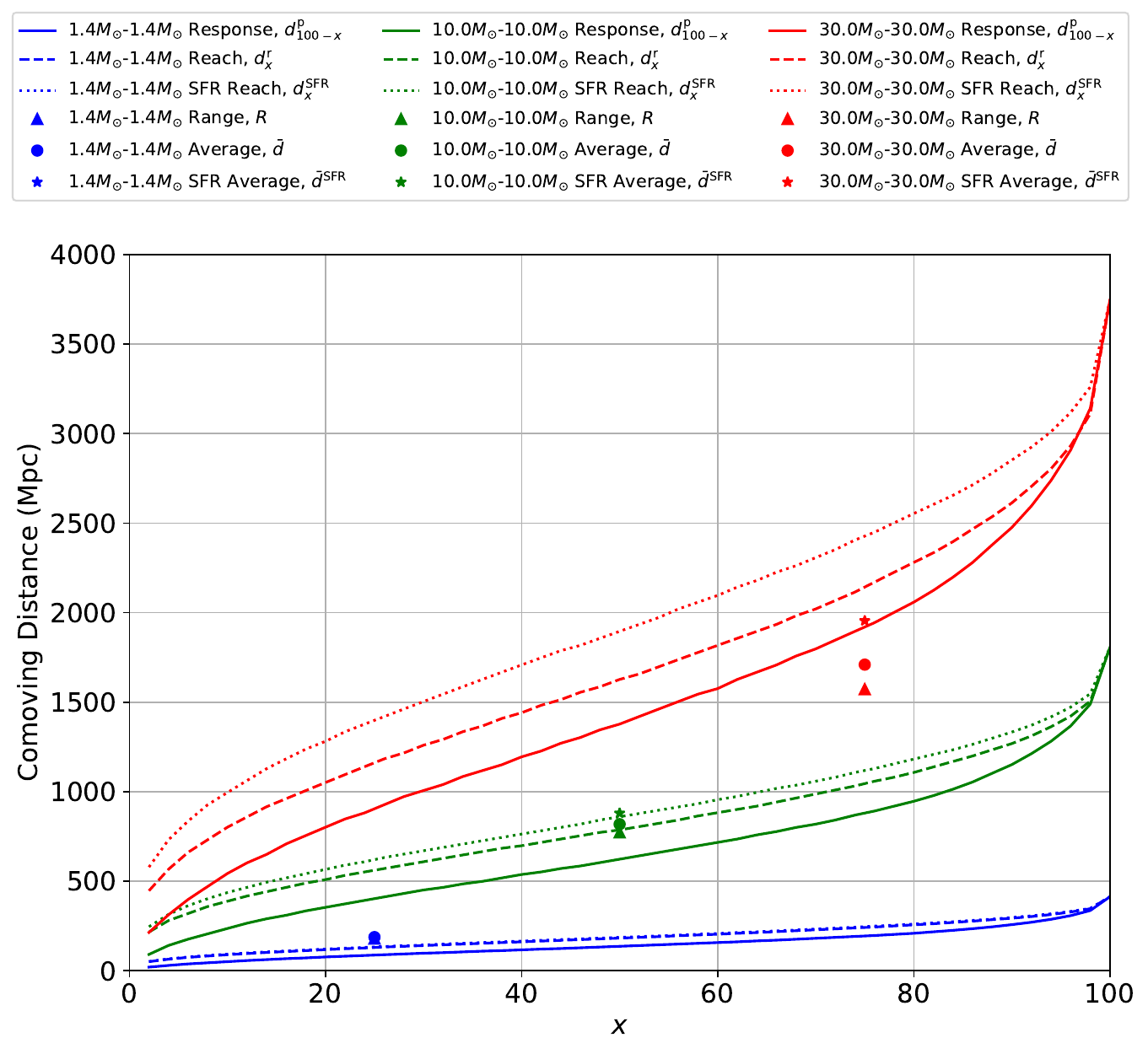}
\caption{\label{fig:2gmeasure}
Distance measures for the sensitivity of the 2nd generation instrument (2G), corresponding to
the Advanced LIGO O4 sensitivity curve in ~\citet{osd}.
{
{The solid line shows the comoving distance at which $(100-x)\%$ of the sources would be detected.} 
The dashed line shows the comoving distance within which $x\%$ of the total detections would take place, and 
the dotted line shows the same quantities but scales the source frame rate density by the star formation rate. 
The triangle, circle, and star are the range, average distance and SFR average distance respectively (see descriptions in Section~\ref{sec:def}).}
}
\end{figure*}

\begin{figure*}
\centering
\includegraphics[width=\textwidth]{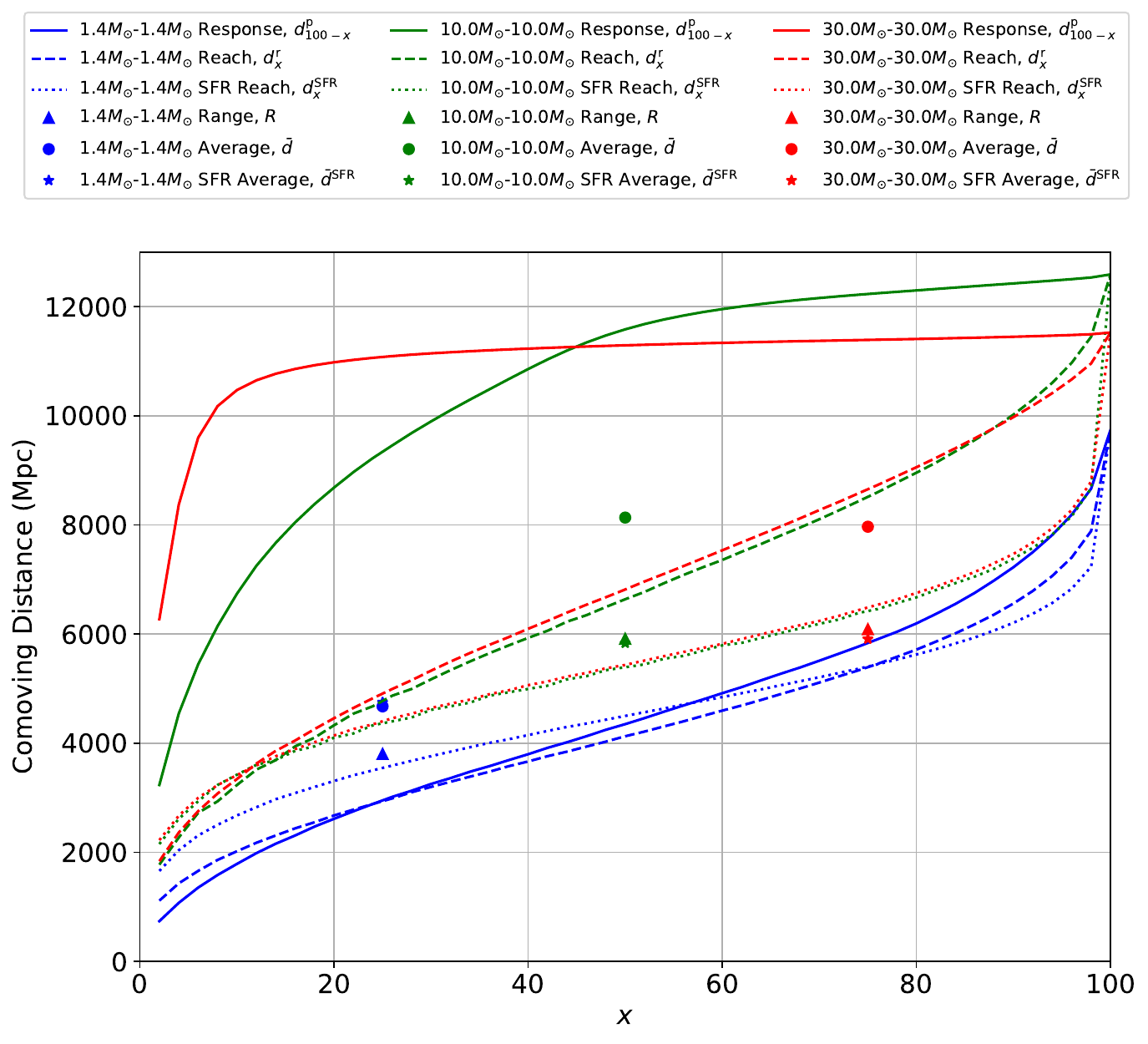}
\caption{\label{fig:3gmeasure}
Distance measures for one of the proposed 3rd generation instruments (3G),
corresponding to the ``CE2'' curve of ~\citet{ce}.
{
{The solid line shows the comoving distance at which $(100-x)\%$ of the sources would be detected.} 
The dashed line shows the comoving distance within which $x\%$ of the total detections would take place, and 
the dotted line shows the same quantities but scales the source frame rate density by the star formation rate. 
The triangle, circle, and star are the range, average distance and SFR average distance respectively (see descriptions in Section~\ref{sec:def}).}
}
\end{figure*}

In Table~\ref{table:summary} and Figures~\ref{fig:2gmeasure} and~\ref{fig:3gmeasure} we present values for these quantities, for three different classes of compact binary coalescence sources:
$1.4$--$1.4\,M_{\odot}$, $10$--$10\,M_{\odot}$, and
$30$--$30\,M_{\odot}$ (all masses are in the source frame), and for
two different detector sensitivities, 2G and 3G. The 2G curve correspond to the Advanced LIGO O4 curve in ~\citet{osd} (\texttt{aligo\_O4high.txt} file in \url{https://dcc.ligo.org/LIGO-T2000012/public}). 3G corresponds to the ``CE2'' curve of ~\citet{ce}. 
{We choose Advanced LIGO and Cosmic Explorer as the representative sensitivities for the second (2G) and the third (3G) generation detectors. 
There are also other detectors proposing to operate at comparable sensitivities, such as Advanced Virgo~\citep[2G, ][]{2015CQGra..32b4001A}, 
KAGRA~\citep[2G, ][]{2013PhRvD..88d3007A}, and Einstein Telescope~\citep[3G, ][]{Hild:2010id,2012CQGra..29l4013S,2016PhRvL.116m1103A}.}

For the waveform of signals, we use the \texttt{IMRPhenomD}
waveform~\citep{2016PhRvD..93d4007K}
\footnote{\texttt{IMRPhenomD} is a frequency-domain phenomenological model gravitational waveform approximant for the inspiral, merger, and ringdown
of non-precessing and aligned-spin black-hole binaries.}. We assume no spin and ignore the tidal effect.

There are a number of important aspects to the distance quantities:

\begin{enumerate}
  \item  We assume a source is ``detected'' if the SNR of the source
  in a single detector with the given noise curve is $\rho>8$. This
    threshold is arbitrary, and corresponds roughly to an SNR=12 network threshold
    for two {equivalent} detectors; it can be
  trivially generalized to different thresholds and networks of detectors.

\item All of the quantities above include the effects of redshift on the gravitational waveform (see \S\ref{sec:cosmo},{~\citet{1987GReGr..19.1163K,2005ApJ...629...15H}}).
The masses quoted are in the {\em source frame}; the waveform of a {$5$--$5\,M_{\odot}$} binary at $z=1$
is identical to the waveform of a {$10$--$10\,M_{\odot}$} {system} at $z=0$, modulo an overall amplitude scaling.

\item Note that for the sake of definiteness we have assumed the cosmological parameters determined by Planck~\citep{2016A&A...594A..13P}: $\Omega_m=0.3065$, $\Omega_{\lambda}=0.6935$, $\Omega_k=0$, $h=0.679$.
Percent level changes in these quantities lead to percent level changes in the distances being quoted.

\item \cg{The detector's calibration uncertainty that affects the magnitude and phase measurement of the GW signals~\cite{2020CQGra..37v5008S} can add in uncertainties on the 
detector sensitivity and the distance measure. For simplicity, the quantities we report only use the point estimates of the (projected) detector sensitivities.}

\item \cg{Reach and Average distances} need to make assumptions for the source-frame rate
density, which is expected to evolve at high redshift. We consider two possibilities: a {\em constant rate} which assumes that the
source population is not evolving in time \cg{(Reach and Average distances)}, and a {\em SFR rate} which
approximates the rate evolution by the star formation rate \cg{(SFR Reach and SFR Average distances)}.
Since many population synthesis models suggest that the merger rate will roughly 
follow the star formation rate{~\citep[e.g.][]{2015ApJ...806..263D}}, we consider a scenario where the merger rate
  directly tracks the shape of the cosmic star formation rate, as represented by
  Eq.~15 of~\cite{2014ARA&A..52..415M}.

\item \cg{Reach, SFR Reach, Average, and SFR Average distances} incorporate {\em time
dilation} by including a $1/(1+z)$ redshift factor to convert the fixed source-frame rate
to a detected rate.

\item The Redshifted Volume quantity assumes that the sources have a constant source-frame rate density and includes
the effect of time dilation and redshift of the waveform.


\end{enumerate}
These aspects are discussed in significantly more detail in Sec.~\ref{sec:howto}.

\section{\cg{Differences between distance measures}}\label{sec:diff}

The different distance quantities enumerated in the previous section represent different ways to encapsulate the performance of a detector to a given source.

The Horizon Distance gives a clear representation of the farthest possible detection. However, because the detector response patterns are not spherical, this quantity is not representative of the general population. Unlike in the EM case, where a large fraction of sources lie near the maximum distance, in the GW case most sources lie significantly closer than the Horizon.

The Redshifted Volume is useful because it gives immediate intuition for how the detection rate scales with sensitivity. If this quantity doubles, then the expected detection rate doubles as well (assuming a constant source-frame rate density).

The Response is useful if one is interested in characterizing a GW detector independent of any assumptions about the
intrinsic rates of the source. This quantity summarizes the {impact of the antenna pattern and the overall sensitivity of the detector.}
(see Sec.~\ref{sec:antenna}).

If one is interested in the median or average distance to which a population of binaries might be detected,
then the Reach distances are more appropriate. These numbers depend on assumptions for the source-frame rate density. 
The ``default'' assumption is that the population follows a constant comoving rate density. Assuming that the populations 
roughly follow the star formation rate, the constant source-frame rate assumption is likely to {underestimate the true rate by factors of a few for $1 <z <3$,}
and {overestimate the rate for} $z\gtrsim3$ (where the true rate density may drop to 0).

To summarize, detectors at Advanced LIGO sensitivity (2G) would find median luminosity distances ($d^{\rm r}_{50}$)
for a detected population of 1.4--1.4$\,\msun$ and 30--30$\,\msun$ binary coalescences of 188 and {2,276}$\,$Mpc, respectively.
For 3rd generation detectors this increases to 9.5 and 30 Gpc.
Note that in this latter case the distances are large enough that the evolution of the intrinsic rate density may bias
these numbers (generically to higher rate densities at high redshift, and therefore to larger distances).

For 3rd generation detectors, the Response for 10--10$\,M_{\odot}$ and
30--30$\,M_{\odot}$ binary mergers are far beyond the median and the average distances (see Figure~\ref{fig:3gmeasure}).
This indicates that the detector is sensitive to the entire population of sources in the universe, and the detected population is limited by an absence of 
sources at high redshift. In addition, the comoving volume element turns over and begins to decrease at high redshift, further 
decreasing the high-redshift sample. Finally, if the source distribution scales with the star formation rate, the population of 
sources is further reduced at {high redshift ($z\gtrsim2$)} as the star formation rate declines.

We note that at low redshift all of the distance quantities are similar 
{(${R}\sim d^{\rm p}_{50}\sim d^{\rm r}_{50}\sim d^{\rm SFR}_{50}\sim {\bar d} \sim {\bar d}^{\rm SFR}$ c.f.} 1.4--$1.4\msun$ binary coalescences in Fig.~\ref{fig:2gmeasure}), as would be expected since cosmological effects should become negligible. For configurations with sensitivity at higher redshift these quantities begin to diverge, reflecting interesting cosmological aspects of GW detector sensitivity.

When the ``SFR'' quantities diverge from their uniform counterparts (e.g., $d^{\rm SFR}_{90}$ compared with $d^{\rm r}_{90}$ in Fig.~\ref{fig:3gmeasure}), this is an indication that the evolution of the sources could become an important factor.

\section{How to ``cosmologize'' gravitational-wave measurements}
\label{sec:howto}

\subsection{\cg{Starting with Euclidean geometry}}
\label{sec:old_way}

It has been common within the GW community to use terms such as horizon distance, range,
average distance, and sensitive volume. These quantities have generally been
defined assuming that space is Euclidean. This is a good approximation
so long as we are considering nearby sources,
where nearby corresponds to $z\lesssim 0.1$ ($\lesssim400$ Mpc). Although this
applies to binary neutron star mergers throughout the Advanced LIGO-Virgo era, for more massive
systems we can far exceed this distance. It is therefore advisable to update
these quantities so that they properly incorporate cosmology.

In particular, a number of simple scaling relations have come into wide use within the community.
For example, \citet{1994PhRvD..49.2658C} and \citet{2005NJPh....7..204F} approximate the BBH waveforms with an inspiral relation,
characterizing the SNR with a simple expression: {$\mbox{SNR}\propto {\cal M}^{5/6}/D$}\cg{, where ${\cal M}$ is the chirp mass and $D$ is the distance of the binary.} 
In addition, the sky sensitivity is described by:
\begin{equation}
\label{eqn:skyprior}
\begin{split}
&\Omega^{1/2} (\theta,\phi,\iota,\psi)= \\
&(F_+^2(\theta,\phi,\psi)(1+\rm{cos}^2\,\iota)^2+4F^2_{\times}(\theta,\phi,\psi) \rm{cos}^2\,\iota)^{1/2},
\end{split}
\end{equation}
where $(\theta,\phi)$ are the sky locations and $(\iota,\psi)$ are the inclination and orientation of the binary.
$F_+$ and $F_x$ are described in \citet{2009LRR....12....2S} and \citet{2011CQGra..28l5023S}.
With ideal sky location and binary inclination and orientation, we find $\Omega=4$ and the binary can
be observed as far as the horizon.
For a Euclidean universe, the ratio between the horizon and range distances simplifies to the well know value of 2.26~\citep{1993PhRvD..47.2198F}.
We can therefore estimate the range {for} BBH sources of chirp mass $\cal M$ as
\begin{equation}
{R}=2\sqrt \frac{5}{96} \frac{c\,(G{\cal M}c^3)^{5/6}}{\pi^{-2/3}}\times \frac{2}{2.26}\times \sqrt I_7,
\end{equation}
where the sensitivity of the detector is encapsulated {in terms of the moment of the 
interferometer's noise power spectrum $S_h(f)$:}  
$$I_7=\displaystyle\int \frac{f^{-7/3}}{S_h(f)}\,df.$$
However, these simple estimates neglect many important factors: the full inspiral-merger-ringdown waveform, cosmological volume, cosmological redshift, time dilation, and rate density evolution. In what follows we discuss these effects in more detail.

\begin{table*}
\centering
\hspace*{-1.5cm}
\begin{tabular}{lccc}
\toprule
\hline\hline
         & $1.4\,M_{\odot}$--$1.4\,M_{\odot}$ 	& $10\,M_{\odot}$--$10\,M_{\odot}$ 	& $30\,M_{\odot}$--$30\,M_{\odot}$ 	\\ \midrule
\hline
\multicolumn{4}{c}{\textbf{2nd generation (Advanced LIGO O4)}} \\
\multicolumn{4}{c}{($z$, $D_{\rm L}$ in Mpc, $D_{\rm C}$ in Mpc)} \\
\hline
Horizon, $d^{\rm h}$ &( 0.1 ,  449.4 ,  410.4 ) &(  0.46 ,  2617.7 ,  1796.6 ) &(  1.14 ,  8014.8 ,  3737.7 ) \\
Redshifted Volume, $V_z$ (Gpc$^3$) & 0.024  & 1.931  & 16.294  \\
Range, $R$ (Mpc) & 179.9  & 772.6  & 1572.7  \\
Response, $d^{\rm p}_{50}$ &( 0.03 ,  139.6 ,  135.4 ) &( 0.15 ,  708.6 ,  618.8 ) &( 0.34 ,  1832.5 ,  1370.0 ) \\
Response, $d^{\rm p}_{10}$ &( 0.06 ,  270.7 ,  255.6 ) &( 0.28 ,  1462.6 ,  1144.7 ) &( 0.66 ,  4103.3 ,  2466.3 ) \\
Reach, $d^{\rm r}_{50}$ &( 0.04 ,  188.3 ,  180.8 ) &(  0.19 ,  926.1 ,  781.5 ) &(  0.41 ,  2276.4 ,  1618.8 ) \\
Reach, $d^{\rm r}_{90}$ &( 0.07 ,  310.8 ,  291.3 ) &(  0.31 ,  1650.9 ,  1261.7 ) &(  0.71 ,  4451.0 ,  2603.7 ) \\
SFR Reach, $d^{\rm SFR}_{50}$ &( 0.04 ,  192.8 ,  184.9 ) &(  0.2 ,  1028.5 ,  854.7 ) &(  0.48 ,  2799.2 ,  1886.9 ) \\
SFR Reach, $d^{\rm SFR}_{90}$ &( 0.07 ,  315.4 ,  295.4 ) &(  0.33 ,  1757.5 ,  1325.8 ) &(  0.79 ,  5096.4 ,  2843.3 ) \\
Average, ${\bar d}$ &( 0.04 ,  194.3 ,  186.3 ) &(  0.19 ,  970.4 ,  813.3 ) &(  0.43 ,  2432.2 ,  1701.3 ) \\
SFR Average, ${\bar d}^{\rm SFR}$ &( 0.04 ,  198.0 ,  189.8 ) &(  0.21 ,  1057.1 ,  874.8 ) &(  0.5 ,  2920.8 ,  1946.0 ) \\
\hline
\multicolumn{4}{c}{\textbf{3rd generation (Cosmic Explorer)}} \\
\multicolumn{4}{c}{($z$, $D_{\rm L}$ in Gpc, $D_{\rm C}$ in Gpc)} \\
\hline
Horizon, $d^{\rm h}$ &( 10.52 ,  112.5 ,  9.8 ) &(  74.8 ,  957.1 ,  12.6 ) &(  29.2 ,  348.8 ,  11.6 ) \\
Redshifted Volume, $V_z$ (Gpc$^3$) & 230.8  & 866.6  & 945.5  \\
Range, $R$ (Gpc) & 3.8  & 5.9  & 6.1  \\
Response, $d^{\rm p}_{50}$ &( 1.42 ,  10.5 ,  4.3 ) &( 30.3 ,  363.2 ,  11.6 ) &( 24.8 ,  292.1 ,  11.3 ) \\
Response, $d^{\rm p}_{10}$ &( 3.84 ,  35.0 ,  7.2 ) &( 62.6 ,  792.9 ,  12.5 ) &( 27.6 ,  327.9 ,  11.5 ) \\
Reach, $d^{\rm r}_{50}$ &( 1.32 ,  9.5 ,  4.1 ) &(  3.2 ,  27.9 ,  6.7 ) &(  3.4 ,  29.8 ,  6.8 ) \\
Reach, $d^{\rm r}_{90}$ &( 3.05 ,  26.6 ,  6.6 ) &(  12.0 ,  130.1 ,  10.0 ) &(  11.7 ,  127.4 ,  10.0 ) \\
SFR Reach, $d^{\rm SFR}_{50}$ &( 1.5 ,  11.2 ,  4.5 ) &(  2.1 ,  16.5 ,  5.4 ) &(  2.1 ,  17.1 ,  5.5 ) \\
SFR Reach, $d^{\rm SFR}_{90}$ &( 2.71 ,  23.0 ,  6.2 ) &(  4.1 ,  37.9 ,  7.4 ) &(  4.2 ,  39.2 ,  7.5 ) \\
Average, ${\bar d}$ &( 1.61 ,  12.2 ,  4.7 ) &(  5.4 ,  51.9 ,  8.1 ) &(  5.0 ,  48.2 ,  8.0 ) \\
SFR Average, ${\bar d}^{\rm SFR}$ &( 1.65 ,  12.6 ,  4.7 ) &(  2.4 ,  19.8 ,  5.8 ) &(  2.5 ,  20.4 ,  5.9 ) \\
\hline\hline
\end{tabular}
\caption{\label{table:summary}
Values for proposed distance measures for different source types.
2G corresponds to the Advanced LIGO O4 sensitivity curve in ~\citet{osd},
and 3G corresponds to the ``CE2'' curve of ~\citet{ce}.
}
\end{table*}

\subsection{{Antenna pattern}}
\label{sec:antenna}

For any set of binary sources at a fixed distance, if you randomly place the sources on
the sky at random inclinations and orientations, you get a distribution of
measured SNR values. This
distribution, when normalized by the maximum SNR in this distribution, is universal; it is
called the {\em antenna pattern}. It does not depend upon
cosmology, the distance, the mass ratio, etc. It also does not depend upon the
noise curve, but it does depend on the type of detector (e.g., 2-arm interferometer),
number of detectors, the detector
orientations, and the relative sensitivities.

This universal single-detector antenna pattern is an incredibly important
tool~\cite{1993ApJ...411L...5C,0264-9381-28-12-125023}. As discussed in more detail
in~\cite{2014ApJ...789..120B,2015arXiv151004615B} and especially the appendix
of~\cite{2015ApJ...806..263D}, we can compress the relevant aspects of the
antenna pattern into a single useful function: the cumulative distribution
function of the antenna pattern, $P(w)$.
Place any compact binary merger at a {\em fixed
distance}, with a random sky position/inclination/orientation, and measure its
SNR in a single GW detector. Let us denote the
maximum possible measured SNR as $\rho_{\rm max}$; this corresponds to a
face-on, overhead binary (see discussion below). We ask: what is the probability
that this binary might have a measured SNR of $\rho$ or greater (where obviously $\rho\leq\rho_{\rm
 max}$)? The answer is given by the cumulative antenna pattern, $P(w)$, with $w=\rho/\rho_{\rm
 max}$. A table allowing for simple interpolation of $P(w)$ can be found at \url{https://github.com/hsinyuc/distancetool}.

Note that if the inclination for all the binaries is fixed (e.g., face-on), and you marginalize over
all sky positions, you get a different distribution. This is still universal, in
the sense that it is independent of cosmology, distance, etc.

Note that if you fix the sky position, but marginalize over all inclinations
(e.g., relevant if the antenna pattern of the combined network is
spherical), then you get yet a different (still universal) distribution.

In what follows we consider the general case, where the sources are not all
at a fixed distance. However, the functional form for $P(>w)$ remains identical,
which is a great simplification.

\subsection{\cg{Local Universe}}
\label{sec:euclid}

For $z\lesssim0.1$, the Universe is well described by Euclidean geometry. In
this case we can define the following quantities:\\

\noindent{\bf Horizon Distance}\/
As discussed briefly in Sec.~\ref{sec:1}, we consider a single GW detector with a known noise curve. For any given binary
coalescence, we define the horizon distance, $d^{\rm h}$ as the maximum distance for which
this binary would have an SNR in the detector of at least~8. This corresponds to placing the binary directly overhead (along a
line perpendicular to the plane of the detector) and in a face-on configuration
(so that the plane of the binary is parallel to the plane of the detector). Any
binary detected with $\mbox{SNR}\ge8$ must be within this distance. This
  horizon distance depends on the masses and spins of the source, as well as
  the noise curve of the detector.\\

\noindent{\bf Sensitive Volume}\/
Let us assume that we have a uniform rate density of binary coalescence throughout the
Universe (e.g., {$100\,{\rm yr}^{-1}\,{\rm Gpc}^3$}). The binaries are randomly
located and oriented on the sky. We would like to calculate the observable rate of binary
coalescence in our detector. Although we can detect binaries as far as $d^{\rm h}$,
most binaries will be neither face-on nor overhead, so in practice we are not
sensitive to all binaries out to that distance. To calculate the true sensitive
volume we need to integrate over the antenna pattern, and average over all
binary inclinations and orientations.
As discussed in Sec.~\ref{sec:euclid}, the antenna pattern can be described through a cumulative
distribution. This gives the probability that a randomly
located/oriented/inclined source {\em at a given distance}\/ will have a
measured SNR $>8w$, where $0<w<1$.
The sensitive
volume is given by
\begin{equation}
V_{\rm sensitive}=V^{\rm h}/f_p^3,
\label{eq:peanut}
\end{equation}
where $V^{\rm h}={4\pi (d^{\rm h})^3/3}$, and the ``peanut factor'', $f_p$,
converts between the horizon distance and the sensitive volume. In this expression $d^{\rm h}$ is the comoving distance corresponding to the horizon luminosity distance. We use the term
``peanut'' because the shape of the sensitive volume is reminiscent of this
tasty snack~(see for example~\cite{0264-9381-28-12-125023}).
For Euclidean geometry we have $f_p=2.264$~\cite{2015ApJ...806..263D}.
This factor is independent of the noise curve and the mass of the binary,
and is solely a function of the (known) {antenna pattern}.

\subsection{Friedmann-Lema\^itre-Robertson-Walker}
\label{sec:cosmo}

We now generalize these quantities to FLRW cosmologies. Although our approach is
general, when we quote numbers or show plots we will assume a standard LCDM
cosmology with $\Omega_m=0.3065$, $\Omega_\lambda=0.6935$, and $h=0.679$. Percent level
changes in these values lead to percent level changes in the distances.

The SNR calculations for a given waveform are identical when generalizing to
cosmology, with two important caveats: 1. the distance is now a luminosity
distance rather than a Euclidean distance, and 2. the redshifting of the
waveform, and therefore the inferred masses, needs to be taken into account{~\citep{1987GReGr..19.1163K,2005ApJ...629...15H}}. The
sensitive volume definition needs to be generalized to take into account three
cosmological effects: 1. the redshifting of the waveform, 2. the redshifting of
time, and 3. the cosmological distance and volume factors.

The redshifting of the waveform leads to two general approaches: one can
consider a fixed mass in the observer (i.e., LIGO and Virgo) frame, or a fixed mass in the
source frame.
We now consider each cases in turn.

\begin{figure}[tb!]
\centering
\includegraphics[width=1.0\columnwidth]{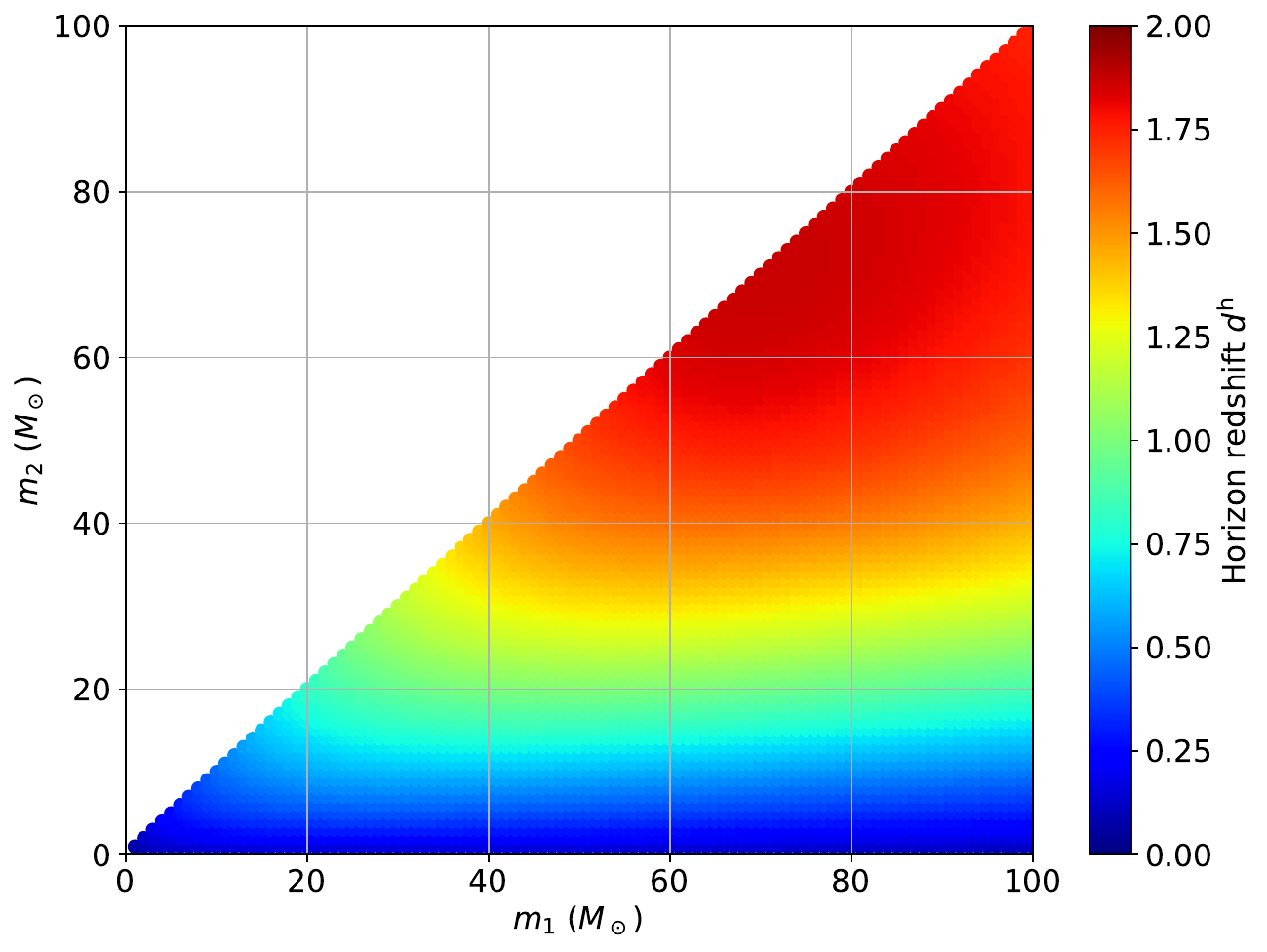}
\caption{\label{fig:source_m_z} Horizon redshift as a function of source frame component masses, assuming
  2G sensitivity. }
\end{figure}

\subsection{Fixed observer frame mass}

In this section we assume that a GW detector is measuring a waveform
corresponding to a binary with component masses $m_1$ and $m_2$,
where these are the observer frame masses.

\noindent{\bf Horizon distance}\/
 We ask how
far a binary with the same {\em observed}\/
  masses could be detected.
In this case the calculation is straightforward.
The horizon distance is defined exactly as in the Euclidean case, but now
the resulting distance is called a luminosity distance. We note that although
the observed total mass is $M$, if the horizon distance for a given binary
corresponds to a redshift of $z^{\rm h}$, then the {\em physical}\/ source frame
mass is actually $M/(1+z^{\rm h})$.

\begin{figure}[tb!]
\centering
\includegraphics[width=1.0\columnwidth]{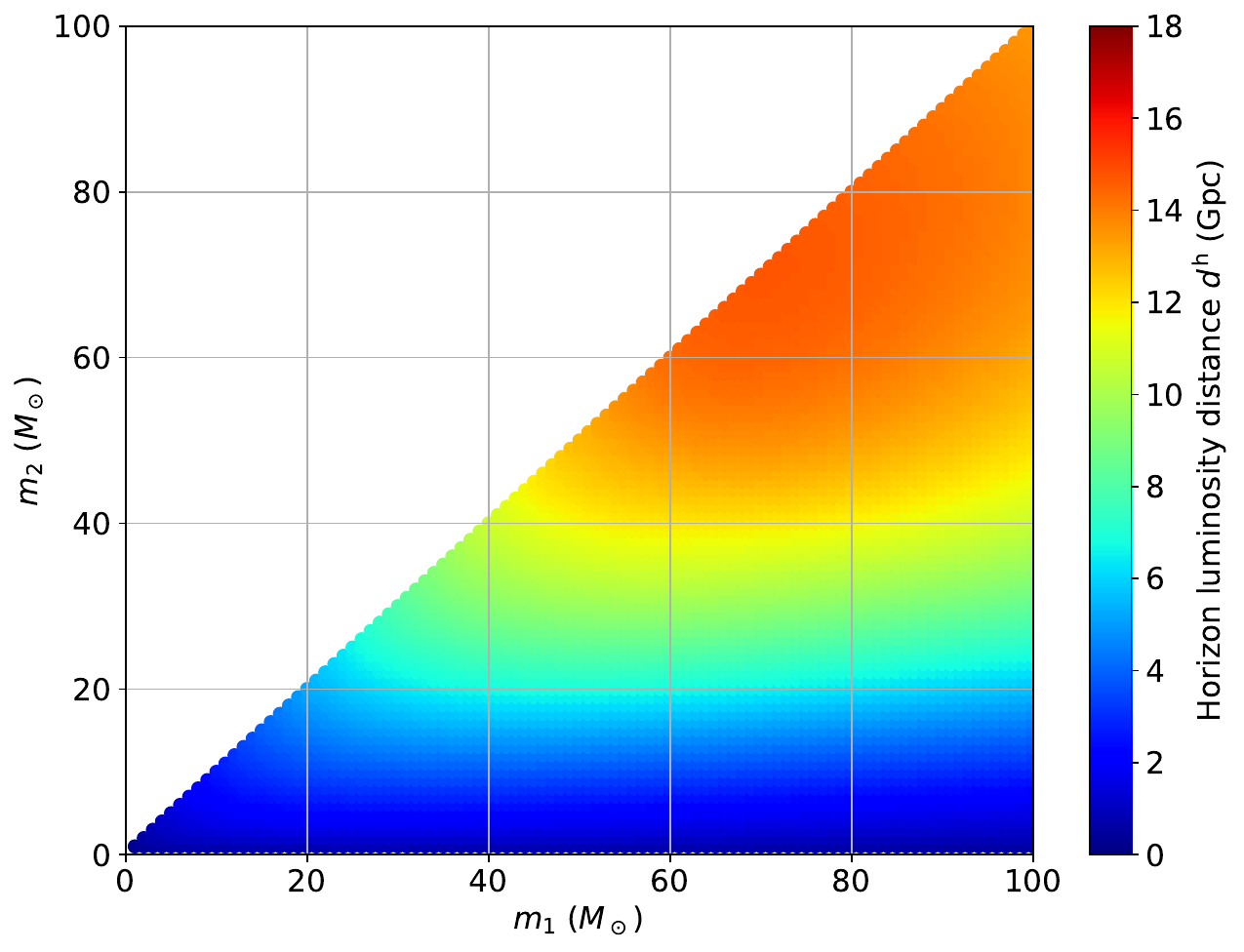}
\caption{\label{fig:source_m_dl} Horizon luminosity distance as a function of source frame component masses, assuming
  2G sensitivity. }
\end{figure}

\noindent{\bf Sensitive volume}\/
This calculation is similar to the
Euclidean case, except that distance becomes luminosity distance and volume
becomes cosmological volume.
Along each line-of-sight one can calculate the luminosity distance at which
$\mbox{SNR}=8$, and we are interested in calculating the volume (in comoving
$\mbox{Mpc}^3$) of this shape.

There are two flavors of sensitive volume, depending on whether one is
interested in estimating a number density or a rate density.
These sensitive volumes are equivalent if the rate of burst sources is fixed in
the {\em observer}\/ frame. For a rate density fixed in the {\em source}\/
frame, the rate density sensitive volume is generally less than the number
density sensitive volume because of redshifting in time of the burst sources.
For example, if we
know that there is 1 (continuous, not burst!) source per comoving
$\mbox{Mpc}^3$, then if we had a number density sensitive volume of
$1\,\mbox{Gpc}^3$ we would be able to detect a total of $1\times10^9$
sources.
However, if we assume burst sources with a constant rate in
the source frame then the detected rate is
impacted by redshifting in time. This is the appropriate case for compact binary coalescence sources, such as the binary mergers detected by LIGO-Virgo thus far. We then define a detection-weighted sensitive volume, or ``redshifted volume'' for short,
so that multiplying this volume by the source frame rate provides the correct
detectable rate:
\begin{equation}\label{eq:vol}
\begin{split}
&V_{\rm sensitive}= \\
& \frac{\displaystyle\int_{D_c<d^{\rm h}(\theta,\phi,\psi,\iota)} \frac{D_c^2}{1+z(D_c)}dD_c\,d\Omega\,\rm{sin}\iota\,d\iota\, d\psi}{\displaystyle\int \rm{sin}\iota\,d\iota\, d\psi},
\end{split}
\end{equation}
where $d^{\rm h}(\theta,\phi,\psi,\iota)$ is the comoving distance for which $\mbox{SNR}=8$ for a binary with orientation
$(\iota,\psi)$ along the sky direction $(\theta,\phi)$.
This redshifted volume is less than the number density
sensitive volume discussed above, since the redshifting in time always reduces
the number of sources detected as one goes to larger distances.

\begin{figure}[tb!]
\centering
\includegraphics[width=1.0\columnwidth]{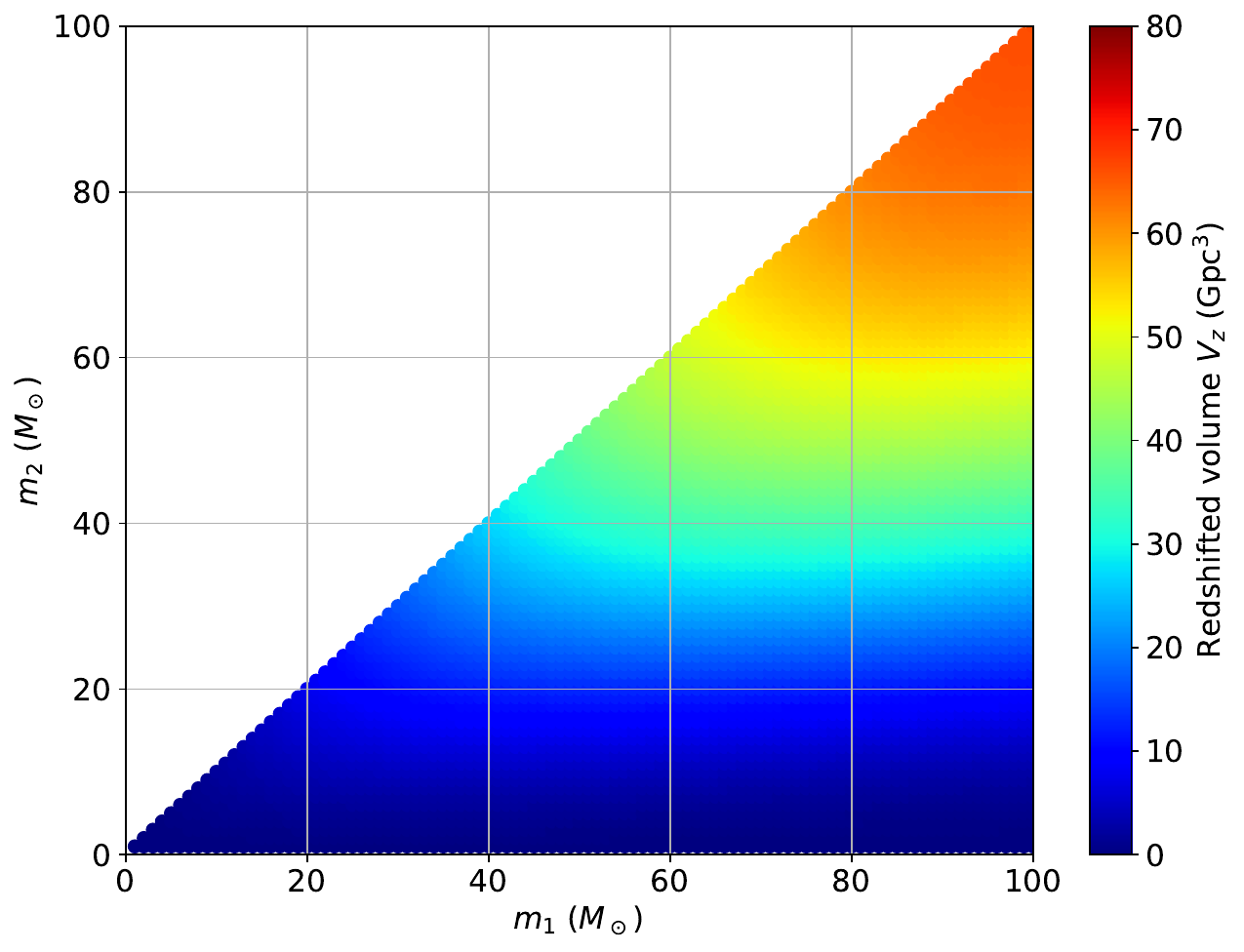}
\caption{\label{fig:source_m_V} Detection-weighted sensitive comoving volume (``redshifted volume'') as
  a function of source frame component masses, assuming 2G sensitivity.}
\end{figure}

\begin{figure}[tb!]
\centering
\includegraphics[width=1.0\columnwidth]{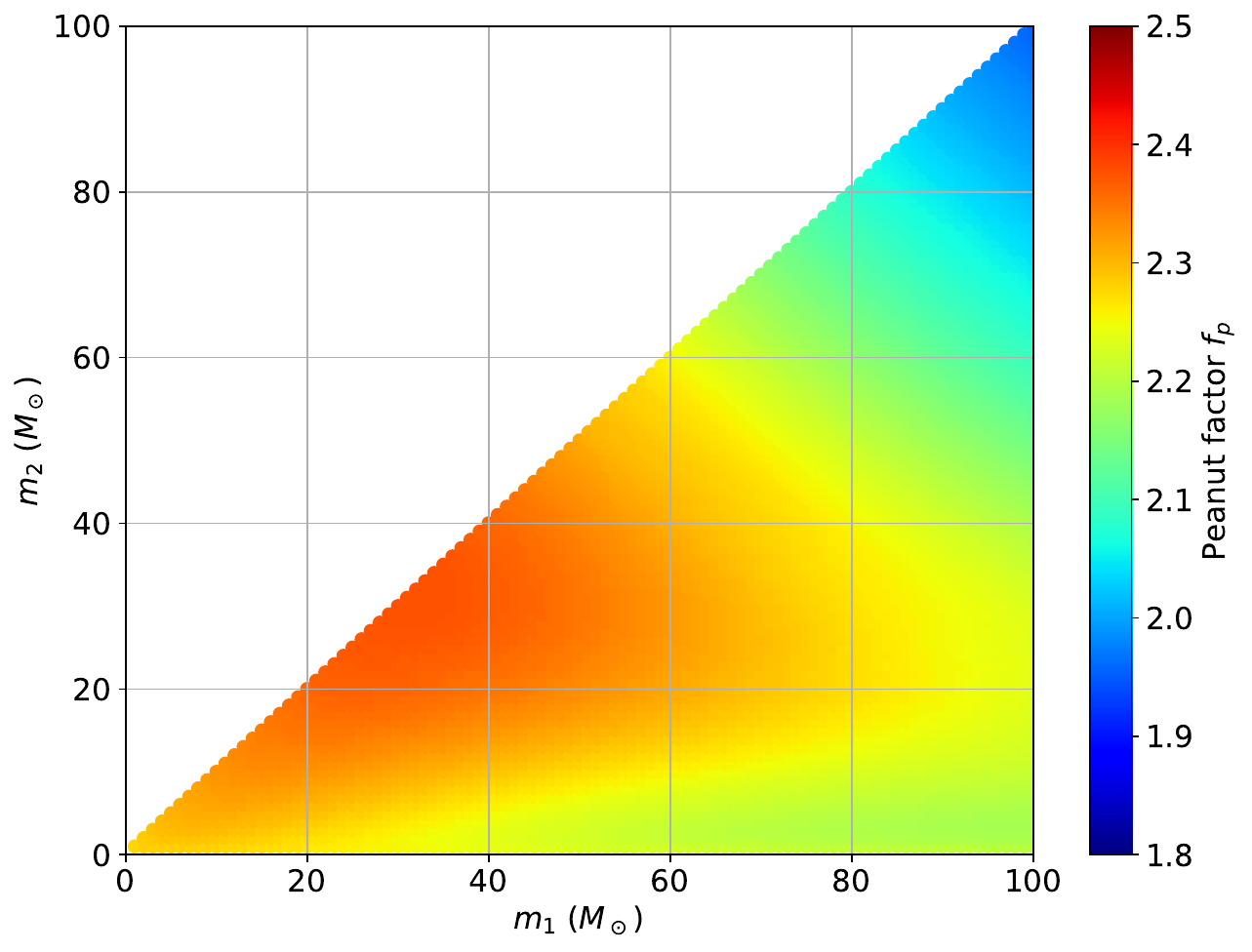}
\caption{\label{fig:source_m_pf} Peanut factor as a function of source frame component masses, assuming
  2G sensitivity.}
\end{figure}

\subsection{Fixed source frame mass}
\label{sec:source_frame}

We now consider the case where the source frame masses are fixed at $M$, and the
observed masses now depend on the redshift of the source: $M_{\rm observed}=(1+z)M$.\\

\noindent{\bf Horizon distance}\/
The horizon distance is given by solving for the distance at which a face-on
overhead binary will be measured with $\mbox{SNR}=8$ for a
binary of mass $(1+z)M$. Values for the horizon redshift as a function of the
source frame component masses are shown in Figure~\ref{fig:source_m_z}. The
equivalent plot for luminosity distance is shown in Figure~\ref{fig:source_m_dl}.
The horizon redshift for GW150914 in the Advanced LIGO O4 sensitivity (2G), keeping the source frame
masses fixed, would be 1.21 (corresponding to a horizon luminosity distance of 8,553 Mpc).
{A web calculator for horizon distance is available at
\url{https://users.rcc.uchicago.edu/~dholz/gwc/}.}

\noindent{\bf Sensitive volume}\/
The sensitive volume is a similar calculation to the fixed observer frame case
above.
However, since we are now considering fixed source frame masses, we are in
effect detecting binaries with different (observer frame) masses at each
distance. This distorts the shape of the sensitive volume, and changes the
values of peanut factor.
In Figure~\ref{fig:source_m_V} we show the redshifted volume as a function of the
source frame masses. The resulting peanut factors are shown in
Figure~\ref{fig:source_m_pf}.
For GW150914, if we fix the mass in the
source frame, we find a sensitive volume of $18\,\mbox{Gpc}^3$ and a peanut
factor of $f_p=2.37$.

\section{Summary}\label{sec:summary}

We have presented a number of quantities to summarize the distance reach of gravitational-wave detectors. In addition to generalizing to luminosity distance and cosmological volumes, and incorporating the {antenna pattern} sensitivity of the detector, we have also incorporated redshifting of the GW waveform, time dilation of the source rate, and possible evolution of the source frame rate density. We present values for a range of binary systems, and a range of detector sensitivities.

\begin{acknowledgements}
We acknowledge Igor Yakushin for providing the web interface and parallelizing the calculator to make it sufficiently fast for interactive usage. 
{We acknowledge Christopher Berry, Thomas Dent, Reed Essick, Stephen Fairhurst, Erik Katsavounidis, Richard O'Shaughnessy, 
Vivien Raymond, Jameson Graef Rollins, Bangalore Sathyaprakash, Leo Singer, Patrick Sutton and Alan Weinstein for valuable discussions and comments.}
HYC was supported by the Black Hole Initiative at Harvard University, which is funded by grants the John Templeton Foundation and the Gordon and Betty Moore Foundation to Harvard University.
HYC and DEH were partially supported by NSF CAREER grant PHY-1151836 and NSF grant PHYS-1708081. They were also supported by the Kavli 
Institute for Cosmological Physics at the University of Chicago through NSF grant PHY-1125897 and an endowment from the Kavli Foundation. 
DEH thanks the Niels Bohr Institute for its hospitality while part of this work was completed, and acknowledges the Kavli Foundation 
and the DNRF for supporting the 2017 Kavli Summer Program. 
{JM, ME, and SV acknowledge the support of the National Science Foundation and the 
LIGO Laboratory. LIGO was constructed by the California Institute of Technology and Massachusetts Institute of Technology with 
funding from the National Science Foundation and operates under cooperative agreement PHY-0757058. JC acknowledges the support of 
NSF grant PHY-1607585. This work was completed in part with resources provided by the University of Chicago Research Computing Center.}

\end{acknowledgements}

\bibliography{bibfile}

\end{document}